\documentstyle[aps,prl,epsf,twocolumn]{revtex} 

\begin{document} \title{ 
Double chains with base pairing: delocalization irrespective to DNA sequencing
}

\author{R.A.Caetano and P.A.Schulz}

\address{Instituto de F\'{\i}sica Gleb Wataghin, UNICAMP, Cx.P.  6165,
 13083-970, Campinas, SP, Brazil}

\date{today}

\begin{abstract}
The question of whether DNA is intrinsically conducting or not is still 
a challenge.
The ongoing debate on DNA molecules as an electronic material has so 
far underestimated a key distinction of the system: the role of 
base pairing  
in opposition to correlations along each chain.
  We show 
that a disordered base paired
double-chain present truly or, at least, effectively delocalized states. This effect is irrespective 
to the sequencing along each chain. 
\noindent PAC(s) numbers: 72.15.Rn,72.80.Le,73.22.-f
\end{abstract}
\maketitle


The convergence of two  
scientific branches -  design of nanometric low dimensional systems in condensed matter 
and controlled 
growth of nucleotide sequences in molecular biology - leads to 
the quest of DNA as a new nanoelectronic material \cite{rmp}. 
 The main question concerning the intrinsic conductivity remains unsolved. 
 Experiments on DNA conductivity
are very controversial:  
metallic\cite{fink}, semiconducting \cite{semiconducting},
insulating\cite{isolante} and even superconducting \cite{kasumov} behaviors have been reported. 
  A recent work, although, has shown that electrons or holes are
responsable for electrical current in DNA \cite{fink} and a biological 
consequence is the fact that a mechanism for 
sensing damaged bases may explore the long range electron migration 
along the molecules \cite{science}. These experiments are very complex due to the
influence of the local environment such as counterions, 
contact resistance, thermal vibrations and even sequence variability, 
which are difficult to control 
in non designed samples
\cite{hao}. 

Here we focus on an intrinsinc factor: the role of 
base pairing in opposition 
to correlations along each chain as a fundamental 
mechanism governing the conductivity. The study of correlations effects 
on disorder has already a long history.
 In summary, disordered one dimensional 
systems should show only stricly localized 
eletronic states \cite{anderson} in absence of electronic interactions.
However, when a short-range correlation is imposed on disorder, 
delocalized states arise, like in 
a random dimmer correlated one-dimentional chain
 \cite{dimer}. On the other hand, DNA can be enginnered to almost any imaginable sequence, therefore a 
completely random sequence is an important limiting case. Some authors claim that 
long range correlations could be present in some genes \cite{peng,holste} but a further claim that such 
correlations could lead to delocalization need further study \cite{rocheprl}.

 The theoretical studies of the electronic properties of DNA range from 
 strictly one-dimensional tight-binding chains \cite{nature}\cite{ulloa},
up to involved {\it ab initio} and
density-functional methods \cite{hao} \cite {ch}. Although partially successful, 
there are severe limitations 
in both approaches. One-dimensional chain models deal 
with effective sites instead of a double-chain 
structure. Therefore, the 
base pairing is not  properly taken into account.  
 On the other hand, involved 
density functional calculations, although giving useful insights on environment 
influence \cite{ch}, have to be limited to a reduced number of model DNA 
molecules. Nevertheless, numerous models in the literature are intermediate to both limits, like the charge ladder model \cite{yi} in which two ordered tight-binding chains are considered together with a Coulomb repulsion between bases. Another approach catching the 
double chain character of the DNA molecule focus on short ordered molecules with a backbone 
chain \cite{cuni}, hindering a clear study of the base pairing effect.

In the present work, we show that DNA-like base pairing leads to delocalization in double chains. Two fundamental aspects are discussed. First, there is indeed a true 
localization-delocalization transition for certain parameters ranges. Secondly, we show that at least a very effective delocalization, in the microelectronic length scale, is induced by base pairing for a wide parameter range that is compatible with the DNA electronic structure. Furthermore, it is 
 important to notice that this
 delocalization is irrespective of correlations along the DNA sequencing. 
We use a double-chain nearest-neighbor (intra and inter-chain) hopping 
tight-binding model to describe the
system.  In this approximation, the Hamiltonian can be writen as:\\
\[H=\sum_{i=1}^{N/2}\sum_{j=1}^{2}[\varepsilon_{i,j}|i,j\rangle \langle j,i| +
V|i+1,j\rangle \langle j,i|+ V|i-1,j\rangle \langle j,i|\] \begin{equation}
\label{hamilton} +V'|i,j+1\rangle \langle j,i|\delta_{j,1} +V'|i,j-1\rangle \langle j,i|\delta_{j,2}]
\end{equation} where $\varepsilon_{i,j}$ is the $(i,j)$ site energy, $V$ is the
intra-chain hopping parameter, $V'$ is the inter-chain hopping parameter and 
$N/2$
is the number of base pairs ($N$ is the total number of sites).

The double chains are constituted by four different sites, 
representing the nucleotides A, C, G and T. 
 These sites are randomly assigned in the first chain with 
 equal concentrations on average, while
 the sites of the second chain  follow the DNA
pairing.  It should be noticed that, unlike the costumary single-chain
correlation (intra-chain correlation)\cite{nature} \cite{ulloa}, 
this is a base pairing correlation, which should necessarily be taken 
into account in order to catch the DNA key features. The inset in Fig. 1(a)
shows a particular double-chain configuration. 
The present model is completely general but will be tied to the quest of the electronic properties 
of DNA by means of a proper parametrization. We use the following site energies:
 $\varepsilon_{A}=8.24$ eV,
$\varepsilon_{T}=9.14$ eV, $\varepsilon_{C}=8.87$ eV and $\varepsilon_{G}=7.75$
eV, according to values suggested in the literature \cite{rochereview}. The hopping parameter 
for intra chain adjacent nucleotides is taken to be $V=1.0$ eV. The inter chain coupling,  $V'$, is 
also treated as a hopping integral. Initially, as a model of DNA molecules we consider $V'<V$,
as expected \cite{isolante,macaga}. All important consequences of a wider parameter range will be discussed at the end.
 To avoid spurious effects due
to a particular configuration, an average over 20 double-chains is always 
undertaken.

 The degree of localization of a state is given by
the participation ratio (PR) \cite{rp}, defined, in tight-binding approximation, by:

\begin{equation} PR=\frac{1}{N\sum_{i=1}^{N/2}\sum_{j=1}^{2}|a_{i,j}|^{4}}
\end{equation} where $a_{ij}$ is the wave function amplitude in the $(i,j)$
site.
 The PR is close to zero for localized states for
 $N \rightarrow \infty$ and reaches the maximum value of 2/3 for delocalized 
states in one dimensional
  ordered systems \cite{rp2}.
 
In order to reveal the 
 effect of base pairing, we compare base paired double chains 
 with the uncorrelated counterparts. Fig. 1(a) shows
the density of states (DOS) for coupled double
chains with 1250 base pairs.  
The hopping parameters here are
 $V=1.0$ eV \cite{rochereview} and $V'=0.5$ eV. There is a no qualitative difference 
 between base paired (indicated by an arrow) and uncorrelated  
 double chains: both show the expected van Hove singularities 
 for a double chain, which are not destroyed by the alloy disorder, although the peaks are more pronounced for the base pairing case. 
 On the other hand, the base pairing introduces a dramatic change 
 in the degree of localization of the states, as can be seen in Fig. 1(b). The 
 base paired chains show two bands of effectively
  delocalized states (indicated by arrows) \cite{length}. 
In the limit of $V'=0$ eV the system reduces to two completely random chains. As soon as an inter chain coupling is turned on, together with the base pairing, the delocalization mechanism is induced (as shown by the raise of the PR values). It would be interesting to verify the limit in which the inter chain coupling dominates over the intra chain one, i. e., $V'> V$. 
It is worth mentioning that bands of truly delocalized states appear due to base pairing if the inter chain hopping is greater than the intra chain one. The inset of Fig. 1(b) shows the participion ratio for a base paired and a uncorrelated double chain with N/2 = 2000 base pairs with the following sites energies $\varepsilon_{A}=10.5$ eV,
$\varepsilon_{T}=10.3$ eV, $\varepsilon_{C}=9.5$ eV and $\varepsilon_{G}=9.7$ eV. These values are close to the DNA parameters \cite{rochereview} but chosen to be symmetric relative to a reference level at 10.0 eV. The hopping parameters are:  $V=2.0$ eV and $V'=3.0$ eV. The base paired case shows two bands with the PRs given by smooth curves reaching 2/3, the expected value for completely delocalized states in ordered chains \cite{rp2}. The two peaks with even higher PRs are fingerprints of edge states coupled to the chain continuum.

The degree of localization of an electronic state can be visualized by the 
extension of the corresponding wave functions. We choose a single disorder 
configuration of the system discussed in Fig.1 and look at 
eigenstates with energies near the PR maximum around 10.7 eV. Fig. 2(a) depicts a wave function 
spreaded out along a entire chain of the base paired system; 
while the inset shows a strongly localized state at the same energy for the 
corresponding uncorrelated double-chain.

 Up to now we have focused on the extreme limit of a completely random DNA 
sequencing. For a random Poly(A)-Poly(T) double strand, the delocalization mechanism is even more important. In Fig. 3 we show the PR for base paired (indicated by an arrow) and unpaired Poly(A)-Poly(T) double chain (base paired) N/2 = 3000 base pairs long. The wave function of a typical delocalized state of the base paired system is shown in the inset \cite{length2}. Effective delocalization has been verified up to 5000 base pairs long double chains (not shown here).
  
A crucial test for the  delocalization 
 is the evolution of PR
  with the length of the system. Fig. 4 shows a map of the average maximum PR as a function of double chain length - up to 3000 base pairs - and inter chain hopping for the completely random sequencing limit. The intra chain hopping is fixed at $V = 1.0$ eV.
 Fig. 4(a) illustrates base paired cases, while Fig. 4(b) is a mapping of the double chains in absence of base pairing. The uncoupled chains limit, $V'=0.0$ eV, is stricly identical in both systems, as expected. As soon as the coupling is turned on, the base pairing leads to higher PR values for any value of inter chain hopping. In order to better illustrate this behaviour we mark one of the contour levels as a guide for the eyes. The horizontal lines represent the DNA-like case discussed in Figs. 1 and 2. The contour level crosses the horizontal line for the base paired case, Fig. 4(a), at a chain length 3 times longer than for the uncorrelated case, Fig. 4(b). For stronger inter chain couplings, this 
length ratio - between base paired and uncorrelated chains - becomes continuously higher. This suggests a crossover from localized to truly delocalized systems, as pointed out for the case shown in the inset of Fig. 1(b). Such localization/delocalization transition can be 
verified quantitatively by investigating the participation number (PN) as a function of chain length. \cite{mirlin}

The PN, given by multiplying the PR by the chain length, should increase with the chain length as $PN = N \times PR \propto N^{D_2}$ \cite{mirlin}. In summary, for a localized state $D_2=0$, i.e., PN is independent in respect to $N$. For delocalized states 
$D_2=1$,  in the present case, while $0 < D_2 < 1$ occurs for wave functions that are effectively delocalized, but occupy only a fraction of the system in 
the thermodynamic limit. In Fig. 5 the average maximum s for various double chains are 
depicted as a function of ln(N/2). The dashed curve corresponds to base paired double chain (DNA-like) discussed in Figs. 1 and 2. The continuous line coresponds to the uncorrelated case (also shown in Figs. 1 and 2). For the lengths investigated, we clearly see the localized behavior in the uncorrelated case ( approximately constant PN). 
The base paired case (dashed curve) shows an effectively delocalized character. It is interesting to notice that truly delocalized case, shown in the inset (see also the inset in Fig. 1(b)): the exponential growth in the present representation corresponds to the linear behavior for delocalized states. 

In conclusion, we investigate the effect of base pairing on the 
transport properties of DNA-like systems.
 Delocalization in coupled
one-dimensional chains has been reported only for a very particular case \cite{peter}: an {\it odd number} of {\it uncorrelated} 
chains and at a single critical energy.
In the present work we now impose a base pairing that mimics 
the key feature of DNA
molecules and entire bands of delocalized states arise.  
This delocalization is effectively originated by the 
base pairing, since the system has an even (2) and not an 
odd  number of chains \cite{peter}. On the other hand, a 
localization-delocalization transition has been discussed for double chains, but  
with {\it intra-chain} correlated disorder \cite{ossipov} 
but in {\it absence} of base pairing.
 Both results do not apply to a DNA 
like struture.
In our model the base pairing
induces dramatic effects on the degree of localization: base pairing 
leads to an effective localization-delocalization transition irrespective to the 
sequencing. A recent report 
 presents experimental evidences of delocalization of 
 injected charges in DNA \cite{apl} 
 that could be related to base pairs coupling. The present results suggest that DNA is intrinsically a promising 
 electronic material and the hindrance to DNA nanolectronics is solely of 
 technological nature. Finaly, we show strong evidences that base pairing also leads to 
a bona fide localization delocalization transition at certain parameter ranges.

{(a) Density of States (DOS) of a disordered 
double chain with (dashed) and without (continuous line) 
base pairing. The length of the chain is 1250 
base pairs. Inset: schematic representation of a
double chain with C-G and A-T base pair inter-chain correlation (b) PR for the double-chains in (a):
with (indicated by an arrow) and without base pairing. Inset: The same but for different parameters (see text)}

{ Wave function for an eigenstate near the PR maximum
 in Fig. 1(b)(Double-chain with base pairing) Inset: a localized state for the uncorrelated case at the same energy.}

{PR for a Poly(A)-Poly(T) double chain $N/2 = 3000$ base pairs long. Base paired (indicated by an arrow) and uncorrelated cases. Inset: a delocalized wave function for the base paired chain}  

{ Conture plots of the average PR maximum as a function of the chain length and inter chain coupling. (a) base paired double chains, and (b) double chains without 
base pairing. For details see text.}  

{ Participation number as a function of double chain length: with (dashed curve) and without (continuous curve) base pairing. Same parameters as in Fig. 1. Inset: the same plot for the system in the inset of Fig. 1(b).}


\begin{references}
\bibitem{length}
This chain would already be a competitive nanoconductor, since 1250 base pairs represent a length of $0.4 \mu m$.
\bibitem{length2}
 It should be kept in mind that 3000 base pairs correpond to a length of 1$\mu$m.
\end{references}
\end{document}